\documentstyle[12pt,fullpage]{article}
\def\be{\begin{equation}}
\def\ee{\end{equation}}
\def\beq{\begin{eqnarray}}
\def\eeq{\end{eqnarray}}

\begin{document}
\begin{flushright}
TIFR/TH/00-37 \\
\end{flushright}
\bigskip
\begin{center}
{\bf Basic Constituents of the Visible and Invisible Matter --- A
Microscopic View of the Universe} \\[2cm]
D.P. Roy \\[1cm]
Department of Theoretical Physics \\
Tata Institute of Fundamental Research \\
Homi Bhabha Road, Mumbai - 400 005, India 
\end{center}
\bigskip\bigskip

\begin{center}
{\bf Abstract}
\end{center}
\medskip

One of the greatest achievements of twentieth century physics is the
discovery of a very close link between the microcosm and the
macrocosm.  This follows from the two basic principles of quantum
mechanics and relativity, the uncertainty principle and the mass
energy equivalence, along with the standard big bang model of cosmology.  As we probe deeper into the microcosm we
encounter states of higher mass and energy, which were associated with
the early history of the universe.  Thus discovery of the atomic
nucleus followed by the nuclear particles, quarks \& gluons and
finally the $W$ \& $Z$ bosons have recreated in the laboratory the
forms of matter that abounded in the very early universe.  This has
helped us to trace back the history of the universe to within a
few picoseconds of its creation.  Finally the discovery of the Higgs and
supersymmetric particles will help to solve the mystry of the
invisible matter, which abound throughout the universe today, as
relics of that early history.

\newpage

\noindent \underbar{\bf Introduction}
\medskip

Our concept of the basic constituents of matter has undergone two
revolutionary changes during the twentieth century.  The first was the
Rutherford scattering experiment of 1911, bombarding Alpha 
particles on the Gold atom.  While most of them passed through
straight, occasionally a few were deflected at very large angles.
This was like shooting bullets at a hay stack and finding that
occasionally one would be deflected at a large angle and hit a
bystander or in Rutherford's own words ``deflected back and hit you on
the head''!  This would mean that there is a hard compact object
hiding in the hay stack.  Likewise the Rutherford scattering experiment
showed the atom to consist of a hard compact nucleus, serrounded by a
cloud of electrons.  The nucleus was found later to be made up of
protons and neutrons.

The second was the electron-proton scattering experiment of 1968 at
the Stanford Linear Accelerator Centre, which was awarded the Nobel
Prize in 1990.  This was essentially a repeat of the Rutherford
scattering type experiment, but at a much higher energy.  The result
was also similar as illustrated below.  It was again clear from the
pattern of large angle scattering that the proton is itself made up of
three compact objects called quarks.

\vspace{6cm}
\includegraphics{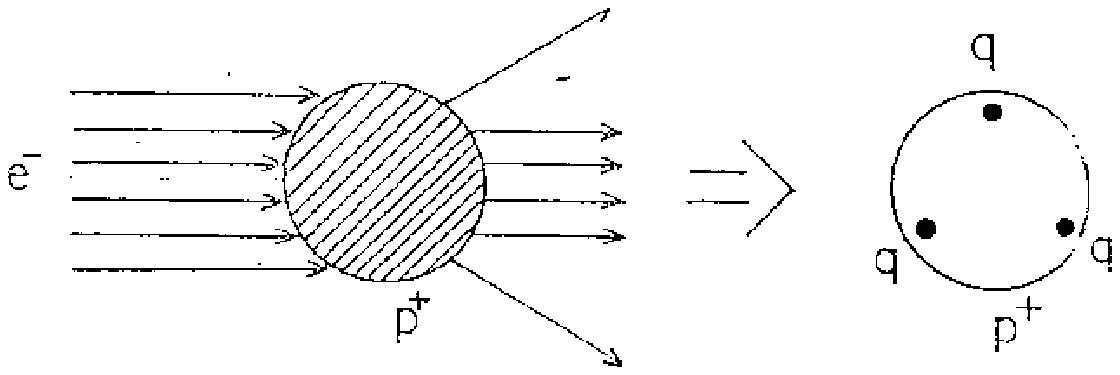}
\label{fig:one.eps}
\begin{center}
Fig. 1.  The SLAC electron-proton scattering experiment, revealing the
quark structure of the proton.
\end{center}

\noindent We now know from many such experiments that the nuclear
particles (proton, neutron and mesons) are all made up of quarks --
i.e. they are all quark atoms.  

The main difference between the two experiments comes from the fact
that, while the dimension of the atom is typically $1 A^\circ =
10^{-10}$ m that of the proton is about $1$ fm (fermi or femtometer)
$= 10^{-15} ~{\rm m}$.  It follows from the famous Uncertainty
Principle of Quantum Mechanics,
\be
\Delta E \cdot \Delta x > \hbar c \sim 0.2 ~{\rm GeV.fm},
\label{one}
\ee 
that the smaller the distance you want to probe the higher must be
the beam energy.  Thus probing inside the proton $(x \ll 1~{\rm fm})$
requires a beam energy $E \gg 1 ~{\rm GeV} (10^9 ~{\rm eV})$, which is
the energy acquired by the electron on passing through a billion
(Gega) volts.  It is this multi-GeV acceleration technology that
accounts for the half a century gap between the two experiments.

It is customary in quantum physics to use the so called natural units,
where one sets both the Plank's constant $\hbar$ and the velocity of
light $c$ equal to unity.  Thus the mass of particle is same as its
rest mass energy $(mc^2)$.  The GeV is commonly used as the basic unit
of mass, energy and momentum.  The proton mass is nearly 1 GeV.
\bigskip

\noindent \underbar{\bf The Standard Model}:
\medskip

As per our present understanding the basic constituents of matter are
a dozen of spin-1/2 particles (in units of $\hbar$) called fermions,
along with their antiparticles.  These are the three pairs of leptons
(electron, muon, tau and their associated neutrinos) and three pairs
of quarks (up, down, strange, charm, bottom and top) as shown below.
The masses of the heaviest members are shown paranthetically in GeV
units.

\begin{center}
Basic Constituents of Matter
\end{center}
\[
\begin{tabular}{|lcccc|}
\hline
&&&& \\
& $\nu_e$ & $\nu_\mu$ & $\nu_\tau$ & 0 \\
leptons &&&& \\
& $e$ & $\mu$ & $\tau(2)$ & -1 \\
&&&& \\
\hline
&&&& \\
& $u$ & $c$ & $t(175)$ & $2/3$ \\
quark &&&& \\
& $d$ & $s$ & $b(5)$ & $-1/3$ \\
&&&& \\
\hline
\end{tabular}
\]

\noindent The members of each pair differ by 1 unit of electric charge
as shown in the last column -- i.e. charge 0 and -1 for the neutrinos
and charged leptons and 2/3 and -1/3 for the upper and lower quarks.
This is relevant for their weak interaction.  Apart from this electric
charge the quarks also possess a new kind of charge called colour
charge.  This is relevant for their strong interaction, which binds
them together inside the nuclear particles.

There are four basic interactions among these particles -- strong,
electromagnetic, weak and gravitational.  Apart from gravitation,
which is too weak to have any perceptible effect,
the other three are all gauge interactions.  They are all mediated by
spin 1 (vector) particles called gauge bosons, whose interactions are
completely specified by the corresponding gauge groups. 
\newpage
\begin{center}
Basic Interactions
\end{center}
\[
\begin{tabular}{|l|lll|}
\hline
&&& \\
Interaction & Strong & EM & Weak \\
&&& \\
Carrier & $g$ & $\gamma$ & $W^\pm \& Z^0$ \\
&&& \\
Gauge Group & $SU(3)$ & \multicolumn{2}{c|}{$\underbrace{U(1)~~ SU(2)}$} \\
&&& \\
\hline
\end{tabular}
\]

\noindent The strong interaction between quarks is mediated by the
exchange of a massless vector boson called gluon.  This is analogous
to the photon, which mediates the electromagnetic interaction between
charged particles (quarks or charged leptons).  The gluon coupling is
proportional to the colour charge just like the photon coupling is
proportional to the electric charge.  The constant of proportionality
for the strong interaction is denoted by $\alpha_s$ in analogy with
the fine structure constant $\alpha$ in the EM case, as shown in
equations (2) and (3) below.  And the theory
of strong interaction is called quantum chromodynamics (QCD) in
analogy with the quantum electrodynamics (QED).  The major difference
of QCD with respect to the QED arises from the nonabelian nature of
its gauge group, $SU(3)$.  This essentially means that unlike the
electric charge the colour charge can take three possible directions
in an abstract space.  These are rather whimsically labelled red, blue
and yelow as illustrated below.  Of course the cancellation of the
colour charges of quarks ensure that the nuclear particles 
composed of them are colour neutral just like the atoms are
electrically neutral.

\vspace{6cm}
\includegraphics{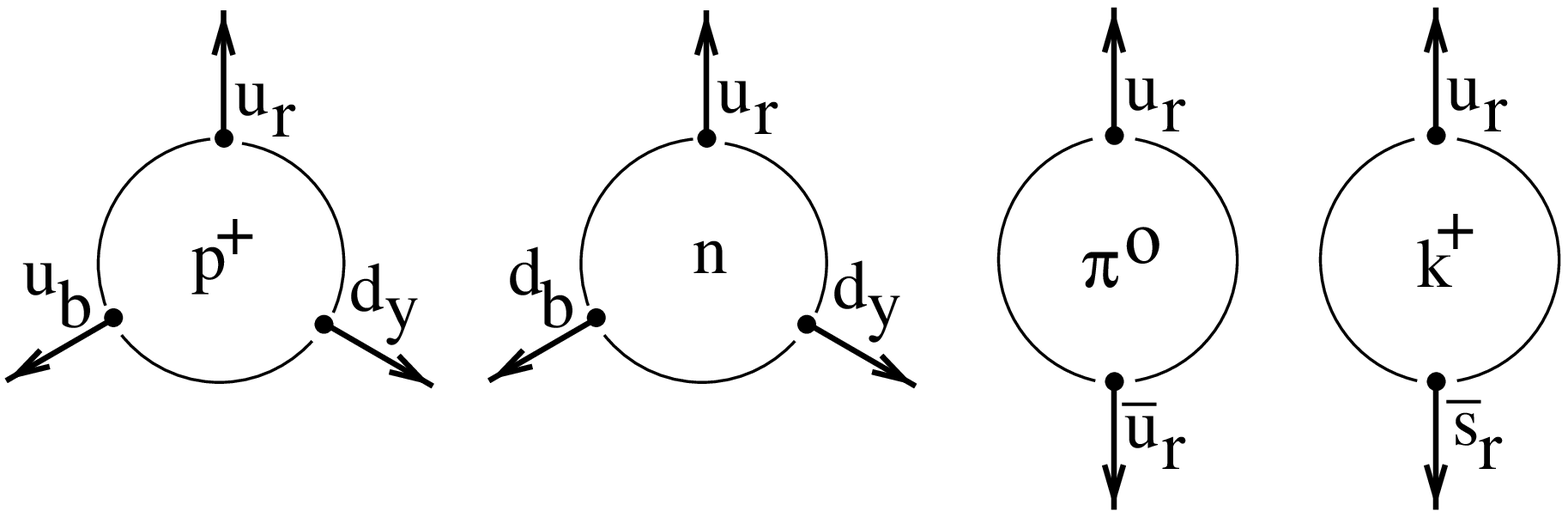}
\label{fig:two.eps}
\begin{center}
Fig. 2. The quark structure of proton, neutron, $\pi$ and $K$ mesons
along with their colour charges (the bar denotes antiparticles and the
subscripts denote colour charge).
\end{center}

A dramatic consequence of the nonabelian nature of the QCD is that the
gluons themselves carry colour charge and hence have self-interaction
unlike the photons, which have no electric charge and hence no
self-interaction.  Because of the gluon self-interaction the colour
lines of forces between the quarks are squeezed into a tube as
illustrated below.

\newpage

\hrule width 0pt
\vspace{6cm}
\includegraphics{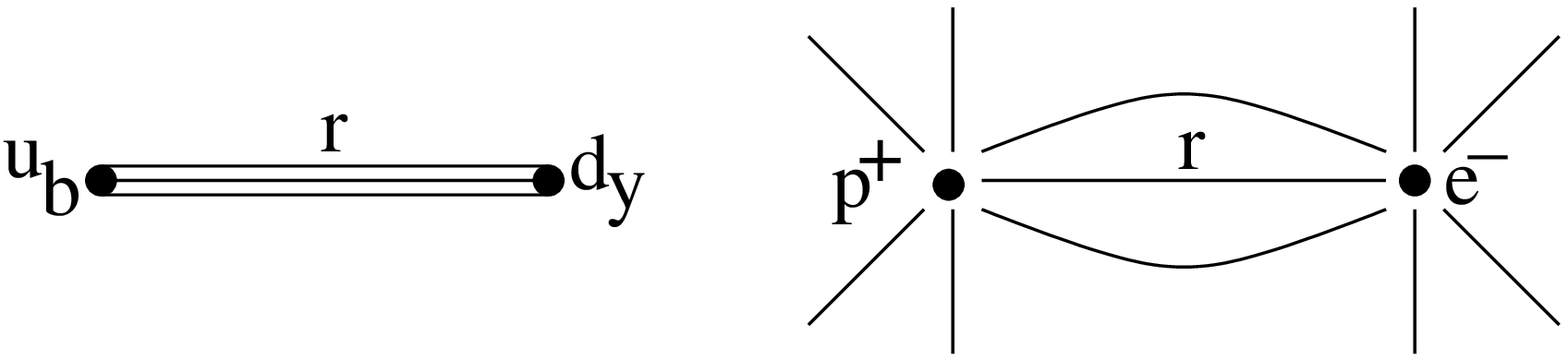}
\label{fig:three.eps}
\begin{center}
Fig. 3. The squeezed (1-dimentional) lines of force between colour charges
contrasted with the isotropic (3-dimentional) lines of force between electric
charges.
\end{center}

\noindent Consequently the number of colour lines of force intercepted
and the resulting force is constant, i.e. the potential increases
linearly with distance
\be
V_S = \alpha_s r.
\label{two}
\ee
Thus the quarks are perpetually confined inside the nuclear particles as it
would cost an infinite ammount of energy to split them apart.  In
contrast the isotropic distribution of the electric lines of force
implies that the number intercepted and hence the resulting force
decreases like $1/r^2$, i.e. the potential
\be
V_E = {\alpha \over r}.
\label{three}
\ee

Finally the weak interaction is mediated by massive vector particles,
the charged $W^\pm$ and the neutral $Z^0$ bosons, which couple to all
the quarks and leptons.  The former couples to each pair of quarks and
leptons listed above with a universal coupling strength $\alpha_W$,
since they all belong to the doublet representation of $SU(2)$
(i.e. carry the same gauge charge).  Because of the mass of the
exchanged particle $M_W$ the weak interaction is restricted to a short
range of $1/M_W$, i.e.
\be
V_W = {\alpha_W \over r} e^{-r M_W}.
\label{four}
\ee
One can understand this easily from the uncertainty principle
(\ref{one}), since the exchange of a massive $W$ boson implies a
tansient energy nonconservation $\Delta E = M_W c^2$, corresponding to
a range $\Delta x = \hbar/M_W c$.  

The weak and the electromagnetic interactions have been successfully
unified into a $SU(2) \times U(1)$ gauge theory.  This is the famous 
electroweak theory of Glashow, Salam and Weinberg for which they were
awarded the 1979 Nobel Prize.  This theory predicts the $W$ and $Z$
boson masses from the relative rates of the weak and the
electromagnetic interactions, i.e. 
\be
M_W = 80 ~{\rm GeV}~{\rm and}~ M_Z = 91 ~{\rm GeV}.
\label{five}
\ee
\bigskip

\noindent \underbar{\bf Discovery of the Fundamental Particles}
\medskip

As mentioned earlier, the up and down quarks are the constituents of
proton and neutron.  Together with the electron they constitute all
the visible matter of the universe.  The heavier quarks and charged leptons
all decay into the lighter ones via weak interaction analogous to the 
nuclear Beta decay.  So they are not freely occurring in nature.  But
they can be produced in laboratory or cosmic ray experiments.  The
muon and the strange quark were discovered in cosmic ray experiments
in the late forties, the latter in the form of $K$ meson.  Next to
come were the neutrinos.  Although practically massless and stable the
neutrinos are hard to detect because they interact only weakly with
matter.  The $\nu_e$ was discovered in atomic reactor experiment in
1956, for which Reines got the Nobel Prize in 1995.  The $\nu_\mu$ was
discovered in the Brookhaven proton synchrotron in 1962, for which
Lederman and Steinberger got the Nobel Prize in 1988.  The first
cosmic ray observation of neutrino came in 1965, when the $\nu_\mu$
was detected in the Kolar Gold Field experiment. 

The rest of the particles have all been discovered during the last 25 years,
thanks to the advent of the electron-positron and the
antiproton-proton colliders.  First came the windfall of the seventies
with a quick succession of discoveries mainly at the $e^+e^-$
colliders: charm quark (1974), Tau lepton (1975), bottom quark (1977)
and the gluon (1979).  This was followed by the discovery of $W$ and
$Z$ bosons (1983) and finally the top quark (1995) at the $\bar pp$
colliders.  Richter and Ting got the Nobel prize for the discovery of
charm quark, while Martin Pearl got it for the Tau Lepton and Carlo
Rubbia for the $W$ and $Z$ bosons. 

\vspace{10cm}

\includegraphics{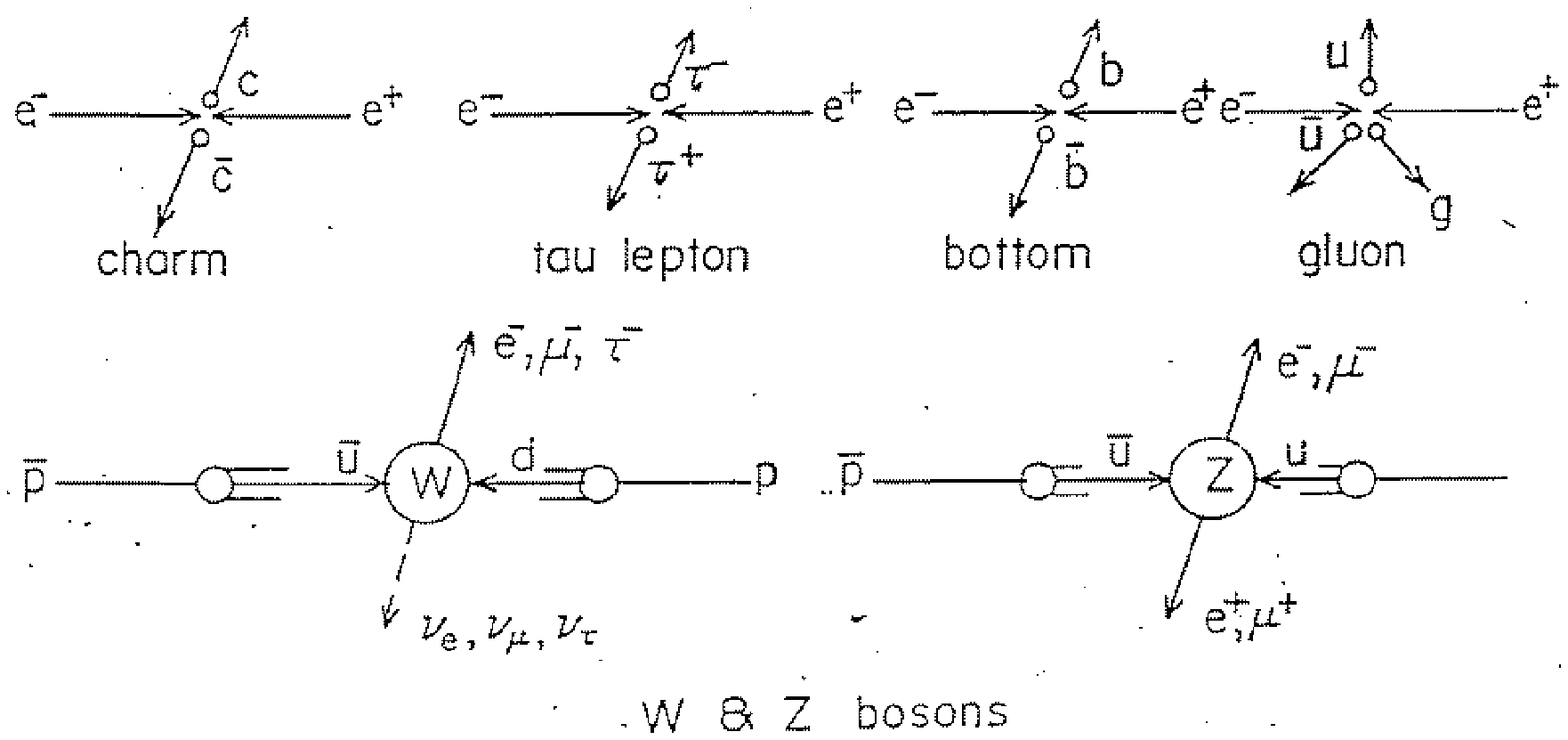}
\label{fig:four.eps}

\begin{center}
Fig. 4. Production of charm quark, tau lepton, bottom quark and gluon
at the $e^+e^-$ collider (upper line) and $W$ and $Z$ bosons at the
$\bar pp$ collider (lower line).
\end{center}

\noindent Fig. 4 illustrates the pair production of charm and bottom
quarks as well as the tau lepton at the electron-positron collider.
The typical lifetime of these particles is of the order of a
picosecond ($10^{-12}$ sec.), corresponding to a range of a few
hundred microns (fraction of a milli meter) at relativistic energies.
Thanks to the high resolution silicon detectors available now, one can
identify these particles before they decay.  On the other hand the $W$
and $Z$ bosons, being very heavy, decay practically at the instant of
their creation.  Nontheless they can be recognised by the unmistakable
imprints they leave behind in their decay products.  The same is true
about top quark production (not shown).

Thus we have seen all the basic constituents of matter and the
carriers of their interactions by now.  But the story is not yet
complete because of the mass problem -- i.e. how to give mass to the
weak gauge bosons (and also matter fermions) without breaking the
gauge symmetry of the Lagrangian.
\bigskip

\noindent \underbar{\bf Mass Problem (Higgs Boson)}:
\medskip

It is a remarkable property of gauge theory that the interaction
dynamics is completely determined by the symmetry of the Lagrangian
under the gauge transformation -- i.e. rotation in some abstract space
defined by the gauge group.  This is similar to the rotational
symmetry of the Lagrangian in ordinary space, leading to the
conservation of angular momentum, except that in this case you assume
the rotation angles to vary from point to point (i.e. be localised).
Then it predicts not only the conservation of gauge charges but also
the presence of massless vector particles (gauge bosons) along with
their couplings to the matter fermions.  But the mass term of the
gauge bosons (as well as those of the matter fermions) can be easily
seen to break the gauge symmetry of the Lagrangian.  On the other hand
the scalar (spin-0) particle masses are symmetric under gauge
transformation.  This can be exploited to give mass to the weak gauge
bosons as well as the matter fermions through the back door --
i.e. they acquire mass by absorbing scalar particles (like a snake
acquiring mass by swallowing a frog).

This is the famous Higgs mechanism.  One assumes a scalar particle of
negative mass square, i.e. imaginary mass.  This leads to a
spontaneous symmetry breaking -- i.e. the ground state of energy is
not symmetric under gauge transformation unlike the Lagrangian.  As a
result the gauge bosons, $W$ and $Z$, can acquire mass.  Besides the
physical scalar particle acquires a real mass, comparable to $W$ and
$Z$ masses.  This is the so called Higgs boson, whose detection will
confirm this mechanism of generating particle masses.  It may be noted
here that one can have a consistent gauge theory of electroweak
interaction in the presence of spontaneous symmetry breaking, since
the Lagrangian retains the gauge symmetry.  This was demonstrated by
t'Hooft and Veltman in the early seventies, for which they got the
Nobel prize last year. 

There are many examples of spontaneous symmetry breaking in physics,
the most familiar one being that of magnetism.  At high temperature
the electron spins of a Ferromagnet are randomly oriented.  But as we
cool it below a critical temperature the electron spins get alligned
with one another because that corresponds to a lower state of energy
(ground state).  Thus while the Lagrangian posseses a rotational
symmetry, this is not shared by the ground state.  The same thing
happens in the Higgs mechanism, except that the rotation is to be done
in an abstract space instead of the ordinary space. 
\newpage

\noindent \underbar{\bf The Standard Model of Cosmology}:
\medskip

A phase transition similar to the Ferromagnetic one had taken place
soon after the big bang, when the universe was only a few
picoseconds ($10^{-12}$ sec.) old.
It was an extremely hot and dense fire ball, with an ambient
temperature of about 100 GeV.  Now the effective mass of a particle
depends on the property of the medium.  And as the universe cooled
down below a critical temperature the effective squared mass of the
scalar particle became negative, resulting in a spontaneous breaking
of the electroweak symmetry.  At this stage the $W$ and $Z$ bosons
acquired mass along with the quarks and the leptons.  The age of
the universe varied inversely as the square of its temperature.  By
the time the universe was about a microsecond ($10^{-6}$ sec.) old,
the ambient temperature had dropped to about 1
GeV and all the $W$ and $Z$ bosons had decayed along with the
heavy quarks and leptons into the light ones.  At this stage a second
(QCD) phase 
transition took place, which confined the light quarks ($u$ and
$d$) into protons and neutrons.  It may be noted here that the ongoing
heavy ion collision experiments are trying to recreate such a
QCD phase transition in the laboratory; and there are already some
early indications of success.

By the time the universe was a few minutes ($10^2$ sec.) old the
temperature had dropped to a few MeV, which is the typical binding
energy of nuclei.  At this stage Helium and other light nuclei were
formed.  After about one lakh $(10^5)$ years the temperature dropped
down to the atomic binding energy level of a few electron volts.  At
this stage the electrons combined with the nuclei to form neutral
atoms.  Thus the matter decoupled from radiation and the universe
became transparent.  After this the matter particles experienced the
gravitational attraction resulting finally in the formation of
galaxies and stars.  The present age of the universe is about 15
billion years and the ambient temperature about 3 K (1 eV = 12000 K).
\bigskip

\noindent \underbar{\bf Hierarchy Problem (Supersymmetry)}:
\medskip

But this is not the end of the story because of the so called
hierarchy problem -- i.e. how to control the Higgs boson mass in the
mass range of $W$ and $Z$ bosons (around a hundred GeV)?  This is
because the scalar particle mass has a quadratically divergent quantum
correction unlike the fermion and gauge boson masses, since it is not
protected by any symmetry.  Thanks to the uncertainty principle, the
vacuum in quantum mechanics is not empty, but it can contain an unlimited amount of energy
and matter.  And the quantum correction represents the effect of this
medium on the mass of a particle.  Being a divergent effect it would
push up the scalar particle mass to infinity (or a very high cutoff
scale of the theory).  So the question is how to keep it down in the
desired mass range of $W$ and $Z$ bosons?

That the scalar mass is not protected by any symmetry was of course
used above to give mass to gauge bosons and fermions via
Higgs mechanism.  The hierarchy problem encountered now is the flip
side of the same coin.  The most attractive solution is to invoke a
protecting symmetry -- i.e. the supersymmetry (SUSY), which is a
symmetry between fermions and bosons.  As per SUSY all the fermions of
the Standard Models have bosonic Superpartners and vice versa.  They
are listed below along with their spins $S$, where the Superpartners are
indicated by tilde.  The presence of the Superpartners ensure
cancellation of the divergent quantum corrections. 
\[
\begin{tabular}{|cc| cc|cc|c|}
\hline
quarks \& leptons & $S$ & Gauge bosons & $S$ & Higgs & $S$ &
$R$-parity \\
\hline
&&&&&& \\
$q,\ell$ & 1/2 & $\gamma,g,W,Z$ & 1 & $h$ & 0 & +1 \\
&&&&&& \\
$\tilde q,\tilde \ell$ & 0 & $\tilde\gamma,\tilde g,\tilde W,\tilde Z$
& 1/2 & $\tilde h$ & 1/2 & -1\\ 
&&&&&& \\
\hline
\end{tabular}
\]

\noindent The standard particles are distinguished from the
supersymmetric ones by a multiplicative quantum number called
$R$-parity, which is $+1$ for the former and $-1$ for the latter.
Thus the supersymmetric particles have to be produced in pair and the
lightest supersymmetric particle (LSP) has to be stable for $R$-parity
conservation.  In most SUSY models the LSP is the photino $\tilde \gamma$,
or in general mixture of $\tilde \gamma$ and $\tilde Z$, with a mass of
again about a hundred GeV.  Thus it interacts only weakly with matter
and hence hard to detect like the neutrino.  Indeed the LSP is the
leading candidate for constituting the invisible or dark matter of the
universe, which will be the last topic of this article. 
\bigskip

\noindent \underbar{\bf The Invisible (Dark) Matter}:
\medskip

There is a large number of indirect evidences suggesting that the bulk
of the matter of the universe is invisible or dark -- i.e. it has
gravitational but no electromagnetic interaction! The strongest
evidence comes from the rotational velocity of the isolated stars or
hydrogen cloud on the outskirts of galaxies.  One can predict this
velocity by balancing the centrifugal acceleration with the
gravitational one,
\be
{v^2_{rot} \over r} = {G_N M(r) \over r^2}, ~{\rm i.e.} ~v_{rot} =
\sqrt{G_N M(r) \over r},
\label{six}
\ee
where $r$ is the distance of the tracer star from the galactic centre
and $M(r)$ is the galactic mass enclosed within this distance.  Fig. 5
shows the rotation curve for a nearby dwarf spiral galaxy M33,
superimposed on the optical image.  If there were no galactic matter
outside the visible disk the rotation velocity curve would have gone
down like $1/\sqrt{r}$.  Instead it continues to rise towards a
constant value, way beyond the visible disk, suggesting that there is
a lot of invisible matter in and around the galaxy.  Similar rotation
curves have been observed for about a thousand galaxies, including our
own.  And they suggest the mass of the invisible matter to be a over
an order of magnitude larger than the mass of visible matter.  

\newpage

\hrule width 0pt
\vspace{8cm}
\includegraphics{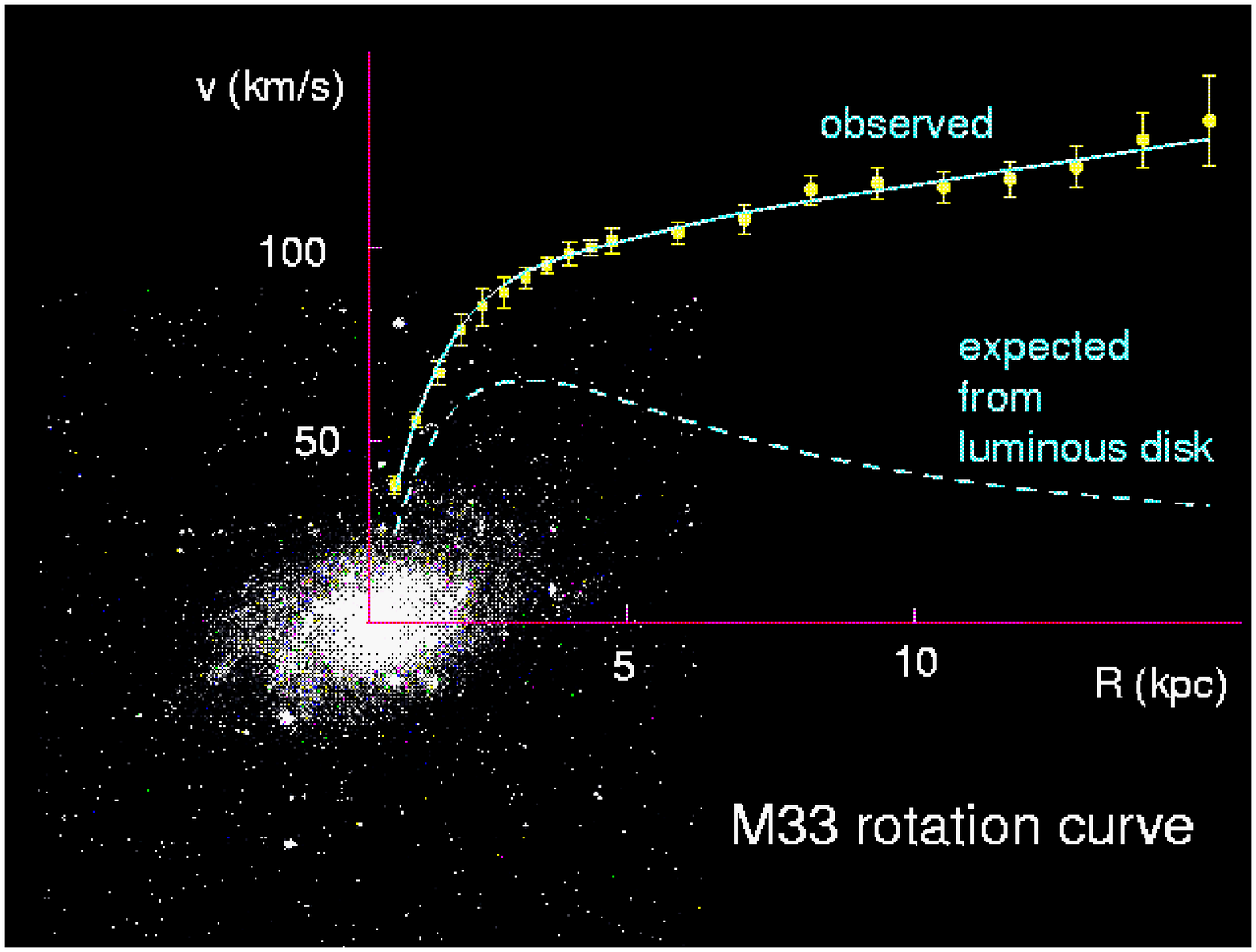}
\label{fig:five.eps}
\begin{center}
Fig. 5.  
observed rotation curve of the nearby dwarf spiral galaxy M33,
superimposed on its optical image (adopted from E. Corbelli and
P. Salucci, astro-ph/9909252, 1999).
\end{center}

The invisible mass could of course come from nonluminous bodies like
small stars or Jupitor, which are more aptly described as dark rather
than invisible matter.  But such objects act as gravitational lense,
which distort the image of distant stars lying behind them by
gravitational bending of light.  And it is clear from these
gravitational lensing experiments that such massive compact
objects can not be a major source of the invisible matter.  The
contribution from gas is also very small.  In principle the neutrinos
could constitute the invisible matter; but their masses are too small
to account for the required density of invisible matter.  The leading
candidate for this is the lightest supersymmetric particles (LSP),
which are weakly interacting like the neutrino but carry a mass of
about a hundred GeV.  These particles were in thermal equillibrium
with $W$, $Z$ and Higgs bosons along with the quarks and leptons when
the universe had a temperature of about a hundred GeV.  But as the
temperature dropped by about an order of magnitude, the density of LSP
came down to a level, where their annihilation rate could not keep up
with the expansion rate of the universe.  The latter took them away
from one another before they could mutually annihilate.  Thus the
total mass of the LSPs at this (freeze out) temperature has remained
the same ever since.  Indeed this scenario predicts the right size of
the invisible matter density we observe today.

Presumably these invisible particles played a pioneering role in
galaxy formation.  Being neutral particles they would have experienced
gravitational attraction long before the formation of neutral atoms,
when the ordinary matter started experiencing this attraction.  Thus
it is quite plausible that the local concentrations of invisible
matter had already formed by the time of formation of the neutral
atoms, to which the latter were gravitationally attracted to form the
galaxies.  Indeed such a scenario of structure formation is supported
by strong observational evidence.  

Thus the invisible matter is a very
important component of the universe.  There are multiprong
experimental efforts for detection of these invisible
particles.  Extremely high precission detectors are being set up in
deep underground experiments to record their interaction with ordinary
matter by measuring the recoil energy of the target nuclei.  A square
kilometer size neutrino telescope is being setup in the polar ice cap
of Antartica to look for the high energy neutrinos coming from the
pair annihilation of these invisible particles, a large concentration
of which is gravitationally trapped at the solar core.  Most
importantly a large proton-proton collider of 27 km circumfurence is
being built by an international collaboration near Geneva to produce
these particles in the laboratory by recreating the extremely high
temperature and energy density, that existed after a nanosecond of the
big bang.  Assuming these invisible particles to be the LSP one can be
reasonably confident of producing them at this collider along with the
Higgs boson.  Thus one hopes to finally solve the mystry of the
invisible matter of the universe as well as to understand how the
visible matter acquires mass.

\end{document}